\documentclass[twocolumn,prb,aps,floats,floatfix,superscriptaddress]{revtex4}

\pdfoutput=1

\usepackage{hyperref}

\usepackage{graphicx}
\usepackage{bm}

\begin{document}


\title{Lattice softening effects at the Mott critical point of Cr-doped V$_2$O$_3$}

\author{S.~Populoh}
\author{P.~Wzietek}%
 \affiliation{Laboratoire de Physique des Solides, Universit\'e Paris-Sud 11, UMR 8502, 91405 Orsay, France}
\email{wzietek@lps.u-psud.fr}
\author{R. Gohier}
\affiliation{Institut des NanoSciences de Paris, Universit\'e Pierre et Marie Curie-Paris 6, CNRS UMR 7588, 140 rue de Lourmel, 75015 Paris, France}

\author{P.~Metcalf}
\affiliation{Department of Materials Engineering, Purdue University, West Lafayette, Indiana 47907, USA}

\date{\today}

\begin{abstract}

We have performed sound velocity measurements in (V$_{1-x}$Cr$_x$)$_2$O$_3$ in the vicinity of the critical point of the first order Mott transition line. The pressure sweeps at constant temperature reveal a large dip in the $c_{33}$ compression modulus, this dip sharpens as the critical point is approached. We do not observe signs of criticality on the shear modulus $c_{44}$ which is consistent with a transition governed by a scalar  order parameter, in accordance with the DMFT description of the transition.
However,  the amplitude of the effect is an order of magnitude smaller than the one obtained from DMFT calculations for a single band Hubbard model.
 We analyze our results using a simple model with the electronic response function obtained from the scaling relations for the conductivity.

\end{abstract}

\maketitle

\section{Introduction}

The Mott transition, which is a metal-insulator transition (MIT) without lattice symmetry breaking, observed in various correlated electron systems, has been one of the major challenges in condensed matter theory
\cite{Imada}.
 The theoretical interest in this phenomenon has been boosted in the recent years by the success of the Dynamic Mean Field Theory (DMFT) approach of  the Hubbard model \cite{Georges96rmp}.
One of the theoretical difficulties concerns the 
choice of the order parameter. Indeed, as the transition does not break any lattice symmetry the order parameter is not clearly identifiable. It was argued that the MIT should be analogous to the liquid-gas transition and belongs to the Ising universality class \cite{Limelette03s}.  Various quantities expressing the ''metallicity'' of the system have been employed to represent an order parameter \cite{Castellani79,Nozieres98epj,Rozenberg99prl,Kotliar00prl}.

Another issue is related to the role of the lattice.
 Tracing back to the earliest ideas, the electron-lattice coupling had often been envisaged as the possible mechanism for a first-order MIT. 
Although the modern DMFT approach can account for the behavior of the resistivity in the vicinity of the critical line within a purely electronic model \cite{Rozenberg99prl}, lattice degrees of freedom do play a role at the transition of real materials, 
where it is accompanied by a discontinuous volume change
 \cite{Kotliar02prl, Majumdar, Merino00prb, Hassan:05PRL}.
 
 On the experimental side, examples of widely studied materials are the Cr doped vanadium oxide \cite{Limelette03s} and organic charge transfer salts $\kappa$-BEDT$_2$X 
 \cite{Lefebvre00prl,Kanoda06jpsj}.
In the latter,   evidences for a lattice softening  near the critical endpoint of the MIT line  were found in ultrasound experiments \cite{Fournier:03PRL} and  by observation of anomalies in the thermal expansion coefficients \cite{Lang}.  

The  Cr-doped vanadium oxide 
(V$_{1-x}$Cr$_x$)$_2$O$_3$ 
is the archetype compound where the MIT can be induced by doping or applying pressure.
Here the transition line ends with a second order critical point near to 
$T=450$ K \cite{McWhan}.  The criticality at this point was studied by electrical transport 
which showed scaling properties expected by DMFT \cite{Limelette03s}. 
 However, experimental facts concerning the role of the lattice in this transition  are rather scarce. 
The information about the lattice response should be important for testing and refining 
 realistic models which take into account the orbital degrees of freedom. 
 Indeed, recent simulations using state-of-the-art LDA+DMFT techniques indicate that in V$_2$O$_3$, the interplay between the correlations and the orbital polarization effects 
modifies substantially the nature of the MIT compared to that of a genuine Mott transition in the one-band Hubbard model \cite{lda-dmft}. 

 In this work we try to shed a new light on the question of the lattice contribution to the MIT through ultrasound measurements in the vicinity of the critical point of (V$_{1-x}$Cr$_x$)$_2$O$_3$.
We show that this experimental approach can address both the questions of the symmetry of the order parameter and its coupling to the lattice degrees of freedom.

\section{Experiments}

All measurements were performed on single crystals with a nominal doping of $x=1.1\%$, prepared using the skull melter technique \cite{Harrison:80MRB}. This Cr concentration ensures that the sample is on the insulating side of the transition at ambient pressure, but a moderate pressure of a few kilobars  drives the system into the metallic state. This doping also corresponds to the one used for transport studies in \cite{Limelette03s}. 

In this study we measured the propagation of acoustic waves along the hexagonal c-axis, for both longitudinal (compression) and transverse (shear) mode. 
The measured sound velocities are proportional to the square root  of the corresponding elastic constants; for the longitudinal mode it is the compression modulus $c_{33}$, and the transverse waves involve the shear modulus $c_{44}$ (the two transverse modes are degenerate for the propagation along the c-axis due to the trigonal symmetry of V$_2$O$_3$) \cite{Truell}.
As it will become clear later, the choice of a high symmetry axis is crucial for the interpretation of our data.  Here we can measure pure compression and shear modes, whereas for almost any other direction of propagation the excited modes would involve a mixture of longitudinal and transverse components of the strain tensor \cite{Truell}.

The crystals were oriented by X-rays, and cut along planes perpendicular to the c-axis to an overall thickness of approx. 2-3~mm. They were then polished on a lapping machine in order to achieve  parallel planes on both sides of the sample and a surface roughness of less than $1~\mu$m (roughly a tenth of the wavelengths used).  The velocity measurements were performed using the standard technique where  
ultrasonic waves are generated by applying  short radiofrequency pulses to a piezoelectric transducer and a phase sensitive detection is used to measure the echo signal \cite{Truell}. We note that  no frequency dependence of the sound velocity was observed within the used range of 200-500 MHz.

One of the technical difficulties encountered is that we cannot use commercial transducers for sound generation and detection, because in high temperature and pressure condition  transducer bonding proves  difficult to achieve and unreliable. Instead, the transducers were directly grown on the sample by sputtering: 
a piezoelectric ZnO transducer was grown on top of a thin chromium bonding layer and a gold contact layer. 
The thickness of the ZnO film was about $10~\mu $m  corresponding to an acoustic resonance frequency in the 400 MHz range. 
Since ZnO crystals grow with their main piezoelectric axis  parallel to the c-axis, in principle only the longitudinal mode should  be excited. Though,  due to imperfections in the ZnO layer it was  possible to excite and observe small transverse mode echoes on some samples.  
The two modes having very different velocities it is in principle easy to unambiguously separate
them when both are excited (thanks to a very broad frequency response of the  transducer
we could do it using very short pulses, less than 100~ns). 
  Very unfortunately, for $T>T_c$  the two velocities are separated by a factor very close to 2 so that the echos of the two modes strongly overlap and it becomes difficult to resolve them even with the shortest available pulses.  Because of this accidental condition we could not obtain the complete set of data for the transverse mode.

The high pressure conditions were obtained using isopentane as pressure liquid and an externally controlled pressure system where the  cell is connected by a capillary tubing to a pressure generator. 
High temperatures are obtained with a small heater inside the cell.
Experiments were performed by sweeping continuously the pressure (at a rate of $\sim$20 bar/min) at constant regulated temperature.
 Our measurements are limited to a pressure of roughly 5 kbar for temperatures in the range of $T\sim 500$ K. These limitations are mainly due to the materials used in the pressure plug and electrical connections.

\section{Results and Discussion}

\begin{figure}[htb!]
\includegraphics[width = 1\linewidth]{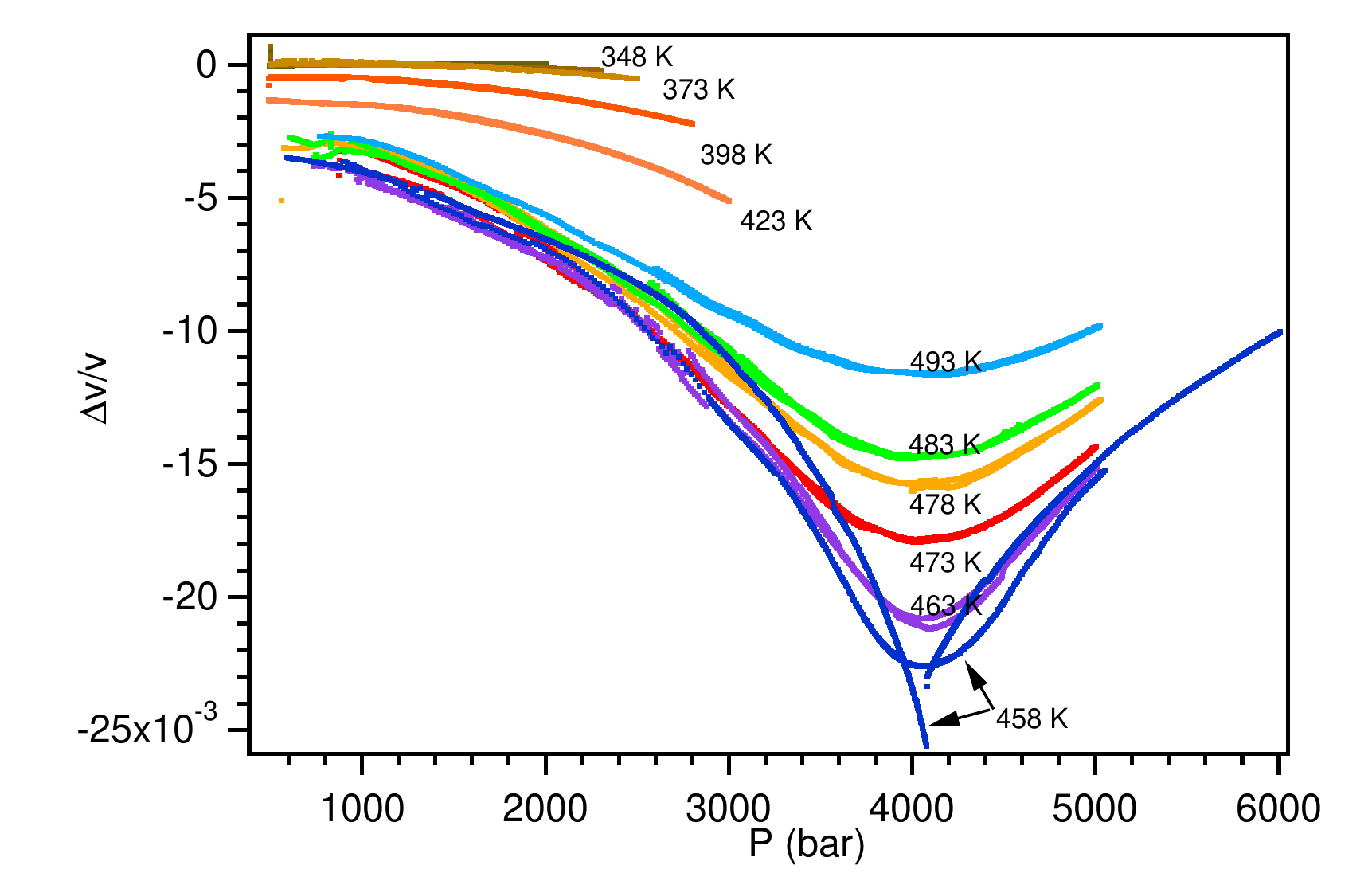}
\caption{Pressure dependence of the relative velocity variation $\Delta v/v$ for the longitudinal mode at various temperatures about the critical point.} 
\label{fig:allT} 
\end{figure}

The evolution of the velocity of the longitudinal sound waves with varying pressure at several fixed temperatures is shown in Fig.~\ref{fig:allT}.
 The  data plotted here have been corrected for the usual linear temperature dependence due to the thermal expansion of the lattice \cite{Varshni}: 
  $\Delta v/v\equiv (v(p,T)-v_{lin}(T))/v_{lin}(T)$ where $v_{lin}(T)$ is a linear function of $T$ representing the temperature dependence of the sound velocity 
 observed far from the critical regime (here for $p\sim 500$ bar in the interval 250-400 K). 
Such a linear temperature dependence has been observed for various elastic constants and it is usually  attributed 
 to anharmonic effects \cite{NicholsYang, Varshni}. 
Therefore Fig.~\ref{fig:allT} represents solely the critical contribution to the sound velocity.
Note that for $T\sim T_c$ the critical regime is spanning a  pressure domain large enough to affect the sound velocity even at ambient pressure, as can be seen from the small negative offset of the curves for temperatures close to $T_c$ in figure \ref{fig:allT}. 
At temperatures where a scan in the full pressure range was taken, the relative sound velocity shows a pronounced minimum around $P_c\sim 4000$ bar.
The largest effect is observed around 458 K which is close to the critical temperature of 457.5 K estimated from transport measurements \cite{Limelette03s}.

The most challenging aspect of these experiments is related to the extreme sensitivity of sound propagation to the sample integrity (we emphasize  that our samples are many times bigger than those used for transport measurements in \cite{Limelette03s}). The sharp volume change at the transition (this change amounts to 1\% at room temperature \cite{jayaraman}) most often gives rise to microfractures resulting probably from releasing some internal stresses frozen during crystal growth. Such fractures spoil or even suppress the acoustic echo signal and eventually make the sample unusable. In practice this means that it is forbidden to cross the first order transition line during pressure or temperature sweeps. 
 For this reason, below the critical temperature pressure scans were only performed in the insulating low pressure region.
 Note also that very close to $T_c$ (below ~470 K), the sharpness of the minimum of $\Delta v/v$ depends on the sample (c.f. two curves for 458 K), 
proving that it is also very sensitive to the crystal quality and 
most likely the differences in the doping inhomogeneity.  
In fact, in samples showing the sharpest curves the micro-fractures often appear even slightly above $T_c$. Therefore in order keep the sample unaltered during the set of pressure sweeps and to get reproducible results we are led to make a compromise and choose a less homogeneous sample which shows a more smooth pressure variation near $T_c$.  
All data shown in figure \ref{fig:allT} were obtained on the same crystal, except one of the curves at 458 K which comes from a sample showing the most pronounced dip. 

\begin{figure}[htb!]
\includegraphics[width = 1\linewidth]{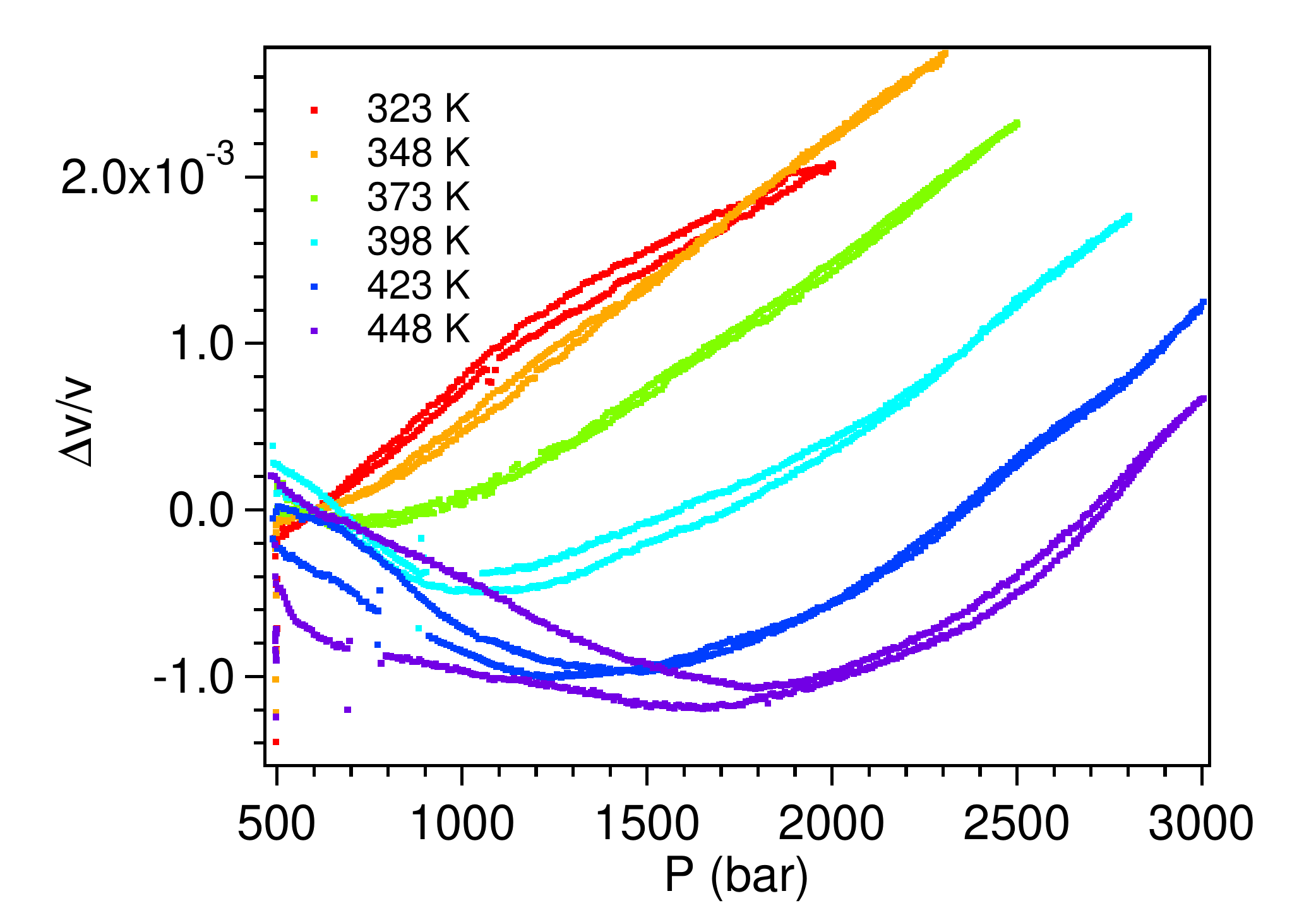}
\caption{Pressure dependence of $\Delta v/v$ of the shear mode at various temperatures}
\label{fig:trans} 
\end{figure}

We now turn to the analysis of the transverse mode velocity,
shown in Fig.\ref{fig:trans}. 
The vertical offsets of different curves were corrected for the temperature dependence by the same procedure as described for the longitudinal mode.
As explained in the previous section, these measurements could only be carried for $T<T_c$ where the transverse echoes are not hidden by the much larger longitudinal echoes, in this region complete scans were not performed to avoid the expected harm to the sample when crossing the first order transition line. 
Nonetheless the data taken up to 448 K show the general trend which is much different that the one observed for the longitudinal mode.
In the region closer to the critical point all curves show a similar slow \emph{upward} shift,  moreover all curves between room temperature and  $T_c$
stay parallel in the region $p\sim$ 2-3 kbar, with no signs of critical behavior at  $T_c$. 
The small increase under pressure is similar to that observed at room temperature in pure V$_2$O$_3$, a compound which does not exhibit the MIT \cite{Nichols:81PRBpressure}. 
The overall behavior of the transverse mode is in contrast to the data of Fig.\ref{fig:allT} where in the same temperature interval a dramatic drop, developing already below 2 kbar, is sharpening as we approach $T_c$. 
While the missing data at the critical point does not allow us to consider this as a rigorous demonstration,  
we think that this finding gives a reasonable evidence that the transverse mode does not exhibit any critical softening close to the metal insulator transition.

For an Ising-class MIT, the absence of criticality of the shear modulus can be inferred 
from simple symmetry arguments. 
 A quite general approach to the effects of a phase transition on the elastic constants was developed by Rehwald \cite{Rehwald}  in the framework of the Landau theory, especially to study soft modes of lattice vibrations observed in structural transitions. In this approach the Landau free energy functional contains coupling terms which represent the interaction 
between the order parameter $\eta$ and the strain tensor $\epsilon$. 
The general expression can be written as a linear combination of 
products of increasing powers of
some components of both the order parameter $\eta$ and the strain tensor $\epsilon$: 
%
$$
	F=\alpha_{ij}\eta_i \epsilon_j + \beta_{ijk}\eta_i\eta_j \epsilon_k + 
	 \gamma_{ijk}\eta_i\epsilon_j \epsilon_k  + \ldots
$$
%
where the  nonzero coefficients $\alpha$, $\beta$, $\gamma$,...  are selected by symmetry.

It has been shown that a singular softening of an elastic constant such as the one in Fig.~\ref{fig:allT} requires the existence of linear coupling between $\eta$ and $\epsilon$.
 The possible combinations can be enumerated by symmetry considerations. In the language of group theory only the components of $\eta$ and $\epsilon$ belonging to the same irreducible representation of the point group symmetry of the disordered phase can be coupled. For a scalar order parameter this means that allowed combinations of the strain tensor elements should form the basis functions of the one dimensional representation $A_{1g}$. In our case, this requirement is fulfilled for the $\epsilon_{zz}$ component  involved in the compression mode, since it is invariant under the operations  of the trigonal symmetry group. On the other hand one cannot construct an invariant from a linear combination of $\epsilon_{zx}$ and $\epsilon_{zy}$, and therefore there is no linear coupling allowed for the transverse mode.
Our failure to detect a critical contribution to the shear mode velocity is therefore consistent with the fact that the Mott transition is expected to be governed by a scalar order parameter.


In the following we will try to quantitatively analyze the data of Fig.\ref{fig:allT}.  
First note that the maximal observed amplitude of the decrease of the relative speed of sound  $\Delta v/v \sim 2\%$ is one order of magnitude lower that the value reported for organic compounds.
It is also smaller by the same factor when compared to the theoretical values given by Hassan \emph{et al.} \cite{Hassan:05PRL} based on DMFT calculations for a single band Hubbard model
(e.g. at $T=1.27\,T_c \approx 580\,\mathrm{K}$   these calculations give a 3\% effect). 
On the other hand there is a qualitative agreement on the pressure dependence, 
namely the 
width of the dip  which is of order of 2 kbar for $T\sim 500\,\mathrm{K}$.

\begin{figure}[htb!]
\includegraphics[width = 1\linewidth]{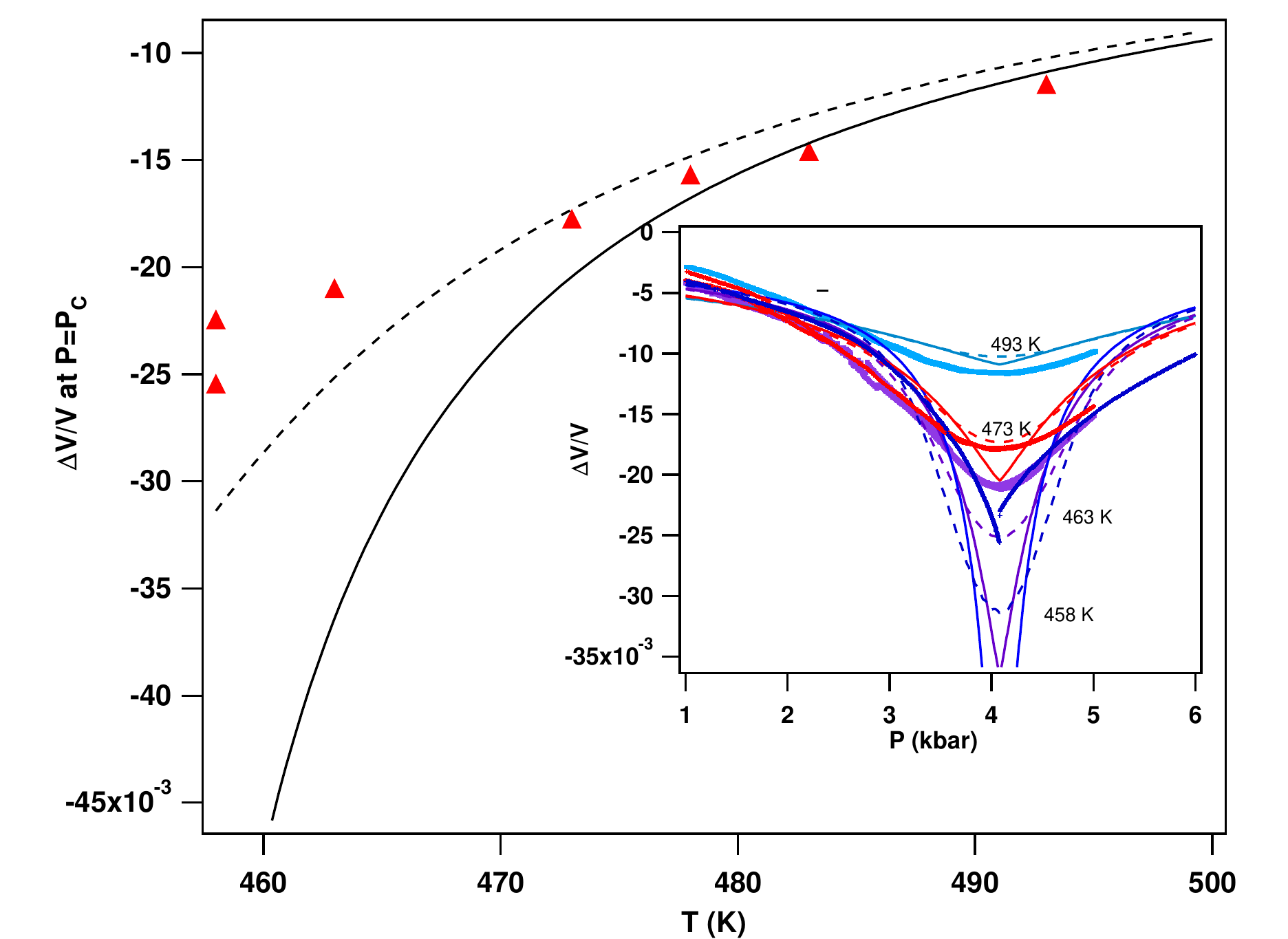}
\caption{Comparison of simulations for $T_{el}=450 K$ (solid and dashed lines) with the experimental data. Main panel: amplitude of the minimum of $\Delta v/v$ in function of temperature; inset: pressure variation for selected temperatures. Dashed lines show results assuming a gaussian distribution of $P_c$}
\label{fig:fit_minimum} 
\end{figure}

Here we will reconsider the same model in a slightly different way.
The compressible Hubbard model is based on the work of Majumdar \emph{et al.} \cite{Majumdar}, 
 who presented the first approach to include lattice effects in this theoretical framework and treated it in the simplest approximation where all phonon excitations are neglected.
They take into account the  dependence of the free energy  on the unit cell volume $V$ through the expression 
$F(V)=F_{e}(D(V))+ E(V)$
where $F_{e}$ is the electronic contribution depending on $V$ through the variation of the bandwidth $D$, and $E(V)$ is the elastic lattice energy. 
The lattice stiffness (compression modulus) 
$$  
K \equiv -V\partial P/\partial V = V\partial^2 F/\partial V^2
$$
can then be expressed as a sum of lattice and electronic contributions:
$K=K_{l}+K_e$.  
The latter is negative and is usually written as
$$
K_e\approx - V\left({\partial D}/{\partial V} \right)^2 \chi_{el}
$$
where the electronic response function is defined as 
$$
\chi_{el} \left(T, D(V)\right) \equiv
-{\partial^2F_{e}}/{\partial D^2}
$$ 
The purely electronic Mott  transition is marked by a divergence of 
 $\chi_{el}$ at $T=T_{el}$, however in the compressible model the instability comes from a divergence of the compressibility $1/K\to\infty$. 
At this point a first order MIT accompanied by a discontinuous volume change occurs to avoid the unphysical region where $K < 0$. 
 Since this happens for a finite value of $\chi_{el}$, the true critical temperature $T_c > T_{el}$. 
 The sound velocity is proportional to $K^{1/2}$ therefore it was predicted that it would vanish as $(T-T_c)^{1/2}$ \cite{Hassan:05PRL}.            
 However, this expansion is derived assuming  $K$  close to zero, i.e. $|K_e|\sim K_l$. Obviously, with $\Delta v/v$ of order of 1\% we are far from reaching such a regime. A more realistic scaling will be obtained assuming $|K_e|<<K_l$ in the  domain studied here. 
 In this limit the sound speed variation obviously follows that of   $\chi_{el}$.
In \cite{Hassan:05PRL} $\chi_{el}$ was determined from DMFT calculations. Here we adopt a different approach using the available experimental data. 
Limelette \emph{et al.} \cite{Limelette03s} argued that in the vicinity of the critical point the excess conductivity in the metallic state $\sigma-\sigma_{crit}$ can be regarded as an order parameter. It was shown that this quantity verifies a universal scaling relation, which, for $T>T_c$ is written as 
$h^{1/\delta}f(ht-\gamma\delta(\delta-1))$
where $t=(T-T_c)/T_c$, $h=(P-P_c)/P_c$, and the exponents were shown to be close to  the mean-field values $\gamma=1$ and $\delta=3$. We have calculated $\chi_{el}=d\sigma/dh$ from the experimental scaling function $f$ given in \cite{Limelette03s}. 
 The apparent symmetry of our experimental curves allows us to suppose that the same scaling in function of $|h|$ can be used on both pressure sides, which is in general expected for $T>T_c$ if $h$ is considered as a conjugate field 
(note that the asymmetry in the DMFT simulations  \cite{Hassan:05PRL} is much less pronounced for $T>T_c$). 
The results of simulations are shown in Figure~\ref{fig:fit_minimum}.
In contrast to the DMFT calculations, we have to treat the amplitude of $\chi_{el}$ as an adjustable parameter. Therefore  $T_{el}$ cannot be  independently determined.   In our simulations we have taken $T_{el}=450 K$  for which the best agreement of the pressure dependence can be obtained at least above 470 K. 
 Our curves go to zero  about 451 K whereas the value in \cite{Hassan:05PRL} gives $T_c-T_{el}\approx 1.4\% \approx 6 $K. This discrepancy comes again from the overestimation of $\Delta v/v$ and $\chi_{el}$ by the same factor. 
Our approach, based on the assumption that conductivity is directly related to $\chi_{el}$, means that we take into account only the effect of the electronic degrees of freedom on the lattice, but neglect the feedback of the lattice on the electronic degrees of freedom. This simplification is now justified by the very small influence of the lattice on the transition temperature.

To check the effect of the sample homogeneity we also calculated the theoretical values of $v$ assuming a gaussian distribution of $P_c$ (dashed curves in Figure~\ref{fig:fit_minimum}). To reach the experimental values of the minimum at 458 K, it is necessary to assume a 20\% gaussian width, an unlikely high value which is clearly incompatible with the sharpest curve at this temperature.  We may speculate that either there is a mechanism precipitating the transition already at very low values of $|K_e|/ K_l$,
or at some point the divergence becomes too narrow to be observed in real samples. 


The pressure dependence of the dip is roughly reproduced by this model but the agreement remains qualitative.
In particular, as we approach $T_c$, the dip on the theoretical curves gets narrow much faster than the one observed experimentally. 
In terms of the scaling function $f(x)$, the pressure profile is the most influenced by the behavior of $f$ at intermediate values of $x$, where the scaling relation is not perfectly obeyed (for these simulations we took  the average scaling function based on the curves reported in  \cite{Limelette03s}).
It is to note however, that independent of the scaling relations, such behavior is expected in our approach as it is reminiscent of the pressure profiles of the conductivity.
At any rate, the failure to reproduce the experimental curves with more accuracy cannot be considered as a major shortcoming for this simplest model where the complex effects of the strain on the multiband structure is parametrized with a simple scalar function $D(V)$.

 In conclusion, we have studied the effects of the electron-lattice coupling in the vicinity of the critical point of the Mott transition line of Cr-doped  V$_2$O$_3$.   Critical effects are only observed for the symmetry-invariant component of the strain tensor,  which strongly supports the Ising scenario for this transition.
  The amplitude of the lattice softening is an order of magnitude smaller than  estimations from a single-band Hubbard  model in the DMFT framework.  
  A simple model where the expression of the electron-lattice interaction is reduced to a simple scalar function $D(V)$ gives a qualitative agreement and is consistent with the scaling relation for resistivity. However a more realistic modeling  would need to  take into account the multi-orbital nature  of this material \cite{lda-dmft}.   

\section{Acknowledgments}

We are indebtful to J.Y.Prieur for his help in setting up the ultrasound experiments. We thank L.~DeMedici, A.~Georges, M.~Marsi and M.~Rozenberg for collaboration on the related issues. We are grateful to H.~Alloul, S.~Brown, H.R.~Krishnamurthy, M.~Poirier and A.-M.~Tremblay  for stimulating discussions. 
This work has been supported by the EC grant n. MEST-CT-2004-514307. 
 

\end{document}